\documentclass[twocolumn,amsmath,nofootinbib,superscriptaddress]{revtex4}
\usepackage{url}
\usepackage{amssymb}
\usepackage{amsfonts}
\usepackage{amsbsy}
\usepackage{graphicx}
\usepackage{hyperref}
\usepackage{array}
\usepackage{multirow}
\usepackage{epsfig}
\usepackage{graphics}
\usepackage{amssymb}
\usepackage{amsmath}
\usepackage{comment}
\usepackage{enumerate}
\usepackage{subcaption}
\usepackage{natbib}
\usepackage{xcolor}
\usepackage{fancyhdr}
\def\spose#1{\hbox to 0pt{#1\hss}}
\def\simlt{\mathrel{\spose{\lower 3pt\hbox{$\mathchar"218$}}
   \raise 2.0pt\hbox{$\mathchar"13C$}}}
\def\simgt{\mathrel{\spose{\lower 3pt\hbox{$\mathchar"218$}}
     \raise 2.0pt\hbox{$\mathchar"13E$}}}
 \def\simpropto{\mathrel{\spose{\lower 3pt\hbox{$\mathchar"218$}}
     \raise 2.0pt\hbox{$\propto$}}}

\def\beq#1{\begin{equation}\label{#1}}
\def\eeq{\end{equation}}
\def\beqa#1{\begin{eqnarray}\label{#1}}
\def\eeqa{\end{eqnarray}}

\usepackage{mathtools}

\fancypagestyle{plain}{
\fancyhf{}
\fancyhead[R]{\fontfamily{cmss}\selectfont
  \fcolorbox{black}{white}{\parbox{1.9in}{\raggedleft\small
    \color{black}DISTRIBUTION STATEMENT A.\\Approved for public release.}}}
}

\parskip=\medskipamount

\begin{document}
\title{Anomaly Detection in Cyber Network Data Using a Cyber Language Approach}
\thanks{This research was developed with funding from the Defense \mbox{Advanced} Research Projects Agency (DARPA). The views, opinions and/or findings expressed are those of the authors and should not be interpreted as representing the official views or policies of the Department of Defense or the U.S. Government.}
\author{Bartley D. Richardson}
\email[]{bartley.richardson@gmail.com}
\affiliation{KeyW Corporation}
\author{Benjamin J. Radford}
\email[]{benjamin.radford@gmail.com}
\affiliation{KeyW Corporation}
\author{Shawn E. Davis}
\email[]{shawn.davis@keywcorp.com}
\affiliation{KeyW Corporation}
\author{Keegan Hines}
\affiliation{Capital One}
\author{David Pekarek}
\affiliation{IronNet Cybersecurity}

\maketitle

% Show fancyhdr on the first page
\thispagestyle{plain}

\date{\today}
\vspace{10mm}

\section{Introduction}

As the amount of cyber data continues to grow, cyber network defenders are faced
with increasing amounts of data they must analyze to ensure the security of
their networks. In addition, new types of attacks are constantly being created
and executed globally. Current rules-based approaches are effective at
characterizing and flagging known attacks, but they typically fail when
presented with a new attack or new types of data. By comparison, unsupervised
machine learning offers distinct advantages by not requiring labeled data to
learn from large amounts of network traffic. In this paper, we present a natural
language-based technique (suffix trees) as applied to cyber anomaly detection.
We illustrate one methodology to generate a language using cyber data features,
and our experimental results illustrate positive preliminary results in applying
this technique to flow-type data. As an underlying assumption to this work, we
make the claim that malicious cyber actors leave observables in the data as they
execute their attacks. This work seeks to identify those artifacts and exploit
them to identify a wide range of cyber attacks without the need for labeled
ground-truth data.

\section{Previous Work}

Previous work has investigated network data for pattern-of-life and anomaly
detection that informs the approaches taken in this paper. In terms of
pattern-of-life, work focuses on identifying and classifying users within a
network \cite{gu2015novel, sharafuddin2010know, verde2014no,abt2014small}.
Clustering for anomaly detection \cite{leung2005unsupervised, portnoy2001intrusion, munz2007traffic} is a common technique for network data due to
the fact that most network datasets are unlabeled and contain no ground truth.
The focus of these works is largely centered around intrusion detection and make
the assumption that intrusions should be anomalous relative to the network as a
whole. Likewise, there exists previous work around the probabilistic suffix tree
(PST) and using it to model and predict protein
families \cite{bejerano2001variations} as well as determine anomalous user
behavior from event logs \cite{liu2013incorporating}.

\section{Data Sources and Experimental Configuration}

\begin{figure}[h!]
\centering
\includegraphics[width=.95\columnwidth]{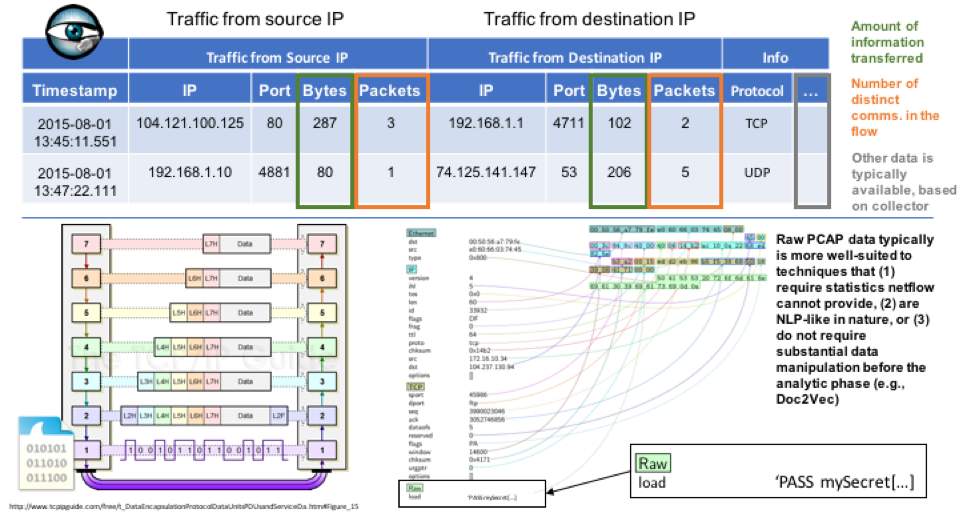}
\caption{Flow-type data (top) compared with PCAP data (bottom). Flow data is aggregated and contains information across multiple PCAP frames, and raw PCAP data provides access to the application payload that may not be represented in aggregated flow data.}
\label{fig1}
\end{figure}

\begin{figure}[h!]
\centering
\includegraphics[width=.95\columnwidth]{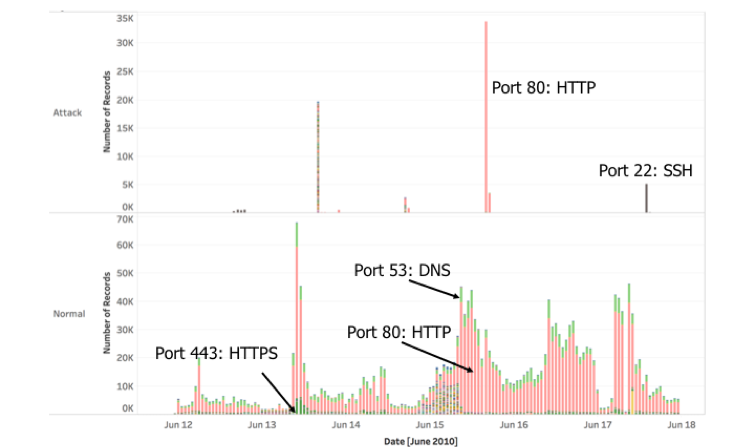}
\caption{Volume of attack (top) and normal (bottom) traffic in the ISCX dataset}
\label{fig2}
\end{figure}

Our data for this effort consists of network traffic data collected in a
traditional compute environment. Specifically, we utilize the University of New
Brunswick Information Security Centre of Excellence (ISCX) Intrusion Detection
Evaluation DataSet \cite{ids2012}. The reason we selected this dataset is that it
contains labeled data for known attacks, and this permits us to produce ROC
curves and calculate AUC in order to evaluate the effectiveness of the
technique. We concentrate on Bro netflow data. By default, Bro creates separate
log files for different actions. For example, the DNS log file contains DNS
resolution requests and associated metadata while the HTTP log file contains,
among other things, GET and POST requests (with the associated URLs) for
activity over HTTP and HTTPS ports. The Bro CONN (connection) log contains
high-level metadata for other log types, including: IP addresses, ports
used, bytes transferred, packets transferred, duration, TCP flags, and protocol.
Bro netflow is aggregated to the session level and is bidirectional, enabling it
to represent many PCAP frames in a single row of information. The relationship
between PCAP and flow-type data is illustrated in Figure \ref{fig1}. The ISCX dataset contains 2,028,053 labeled Netflow
records, with 96.6\% of them labeled normal and the remaining 3.4\% labeled as
attack. The volume of traffic is shown Figure \ref{fig2}. Statistical analysis of real network
traffic is used to create agent-based background traffic in a testbed
environment, and the attacks are real and planned by a white hat team based on
the architecture of the testbed environment.

Our analytics run on a cloud compute environment using Cloudera CDH v5.11.0 and a heavily modified Spark v.2.1.0.cloudera1. The physical
machine contains 6.15TB of addressable RAM with 420 VCores. Actual physical
servers include: 8x24 cores, 22x40 cores, 1x12 core, and 1x4 cores. HDFS is
currently configured for a total capacity of 260.4TiB. Our analytics utilize
Spark MLLib heavily, and input/output is managed via Hive tables in HDFS. It is important
to note that this is a research cloud and is not a production environment.

\section{Creating Sequences of Cyber Data}

\begin{figure}[h!]
\centering
\includegraphics[width=.95\columnwidth]{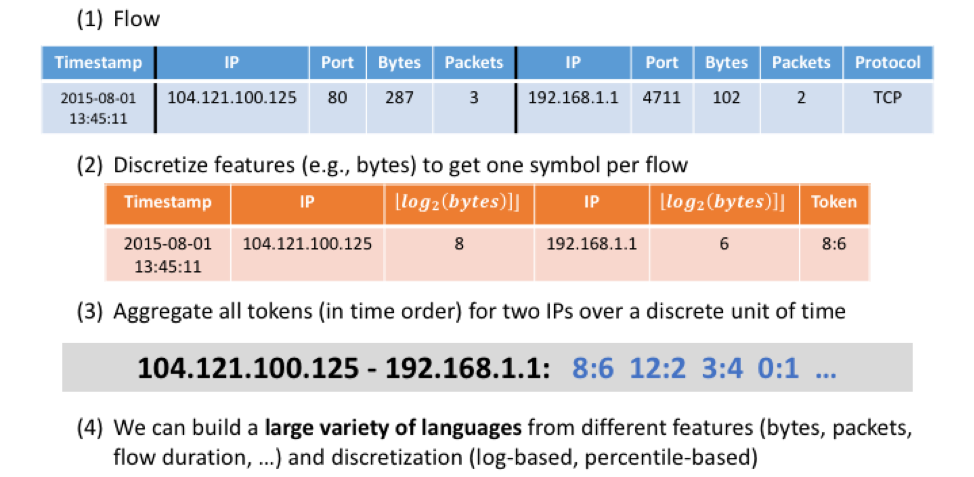}
\caption{Construction of cyber language sequences from flow data}
\label{fig3}
\end{figure}

Before we can apply the analytic to the cyber data, we must transform it into a
sequence of activity (i.e., create the communication language). Figure
\ref{fig3} illustrates this process in detail. In general, some feature
engineering is performed \textit{a priori} to sequence creation. Various combinations of
protocol, port, bytes, packets, and other features can be encoded into discrete
tokens, and sequences of these tokens effectively compress the communication between two
networked computers. For this work, we focus on proto-bytes (a protocol identifier
with the sum of the bytes transferred) and proto-density (a protocol identifier with
the sum of bytes/packet transferred). In order to keep the vocabulary at a
manageable size, we also bin the quantitative features in some way. One method
that has shown promise is to take the floor of the log2 feature value. This allows us to
keep some sense of magnitude (e.g., bytes, KB, MB, GB) while reducing the number
of tokens produced.

Another issue to consider when creating sequences of cyber data is the sequence
length. Sequence length directly relates to time and how long the communication
remains open. Various ways to sessionize exist, including by hour, day, week,
and after 30 minutes of no activity between two IP addresses. For this work, we
construct sequences that terminate after an hour. This has the added benefit of
keeping most sequences to a relatively similar length, so there is not the issue of normalizing all sequences to account for widely varying lengths. These sequences are
created using a parallelized Spark-based approach that can construct multiple
types of sequences relatively quickly.

\section{Modeling Cyber Data using a Probabilistic Suffix Tree}

Creating the PST model is relatively straightforward. The cyber language sequences are fed into a slightly modified PST code model that distributes the learning across a Spark cluster. Typical starting hyperparemeter values set the depth of the tree to 14, the minimum probability to 0.0001, the probability threshold (specifying the minimum probability necessary for inclusion of the suffix in the tree) to 0.0005, and the two smoothing parameters at $\tau=10$ and $\epsilon=0.0$.

\begin{figure}[h!]
\centering
\includegraphics[width=.95\columnwidth]{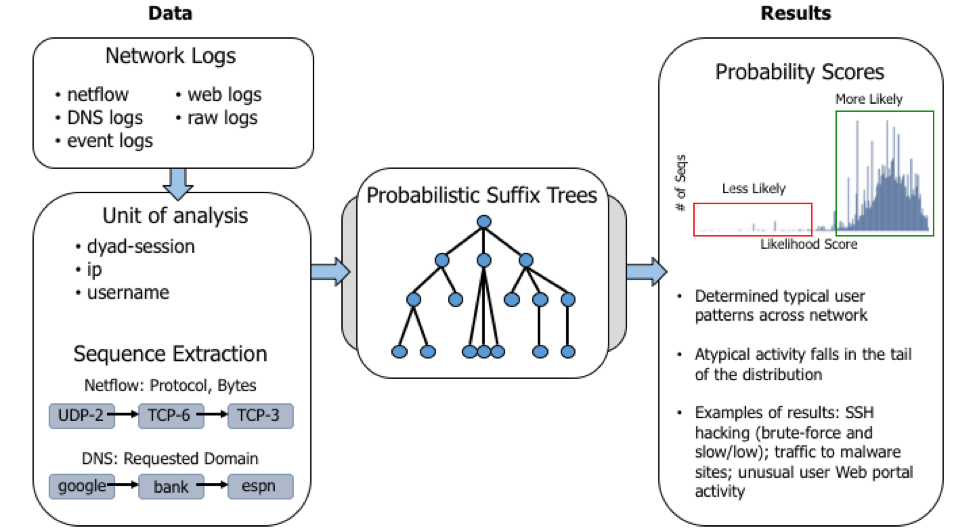}
\caption{Analytic flow for creating PST models from cyber network data}
\label{fig4}
\end{figure}

\begin{figure}[h!]
\centering
\includegraphics[width=.95\columnwidth]{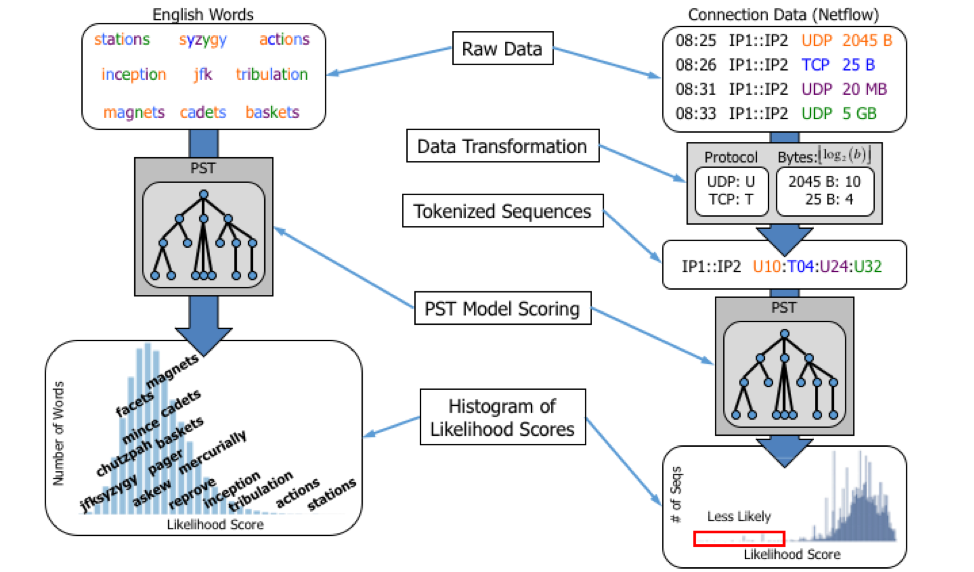}
\caption{Application of PST to natural language (left) and cyber data (right)}
\label{fig5}
\end{figure}

After creating the sequences and the PST model, we then score each sequence
using the model. The overall process is shown in Figure \ref{fig4}. Each sequence receives a likelihood score (a probability between 0 and 1), and we
flag for investigation those sequences that receive a non-zero likelihood score
below a set limit. These sequences represent those less likely to exist in the
data (i.e., anomalous sequences). 
To build intuition, we present
Figure \ref{fig5}. In this figure, the application of PST modeling to English
words is on the left while the cyber application is on the right. In the traditional application, we
seek to quantify a word’s conformity to traditional spelling patterns. Notice
that words like “actions” and “stations” are more likely (therefore further to
the right on the histogram) while words like “chutzpah” and “syzygy” are less
likely (further to the left). In our application to cyber data via construction
of a cyber language, we seek to add interpretability to findings using similar intuition. Our application necessitates
an additional step to transform the data and sequence the tokens. Instead of
analyzing English spelling patterns, we are quantifying spelling patterns of our
tokenized sequences representing machine-to-machine communication. The
underlying assumption is that sequences of less-likely spellings are anomalous
to the network environment where they are observed, and these events warrant
increased scrutiny by a cyber expert.

\section{Experimental Results}

\begin{figure}[h!]
\centering
\includegraphics[width=.95\columnwidth]{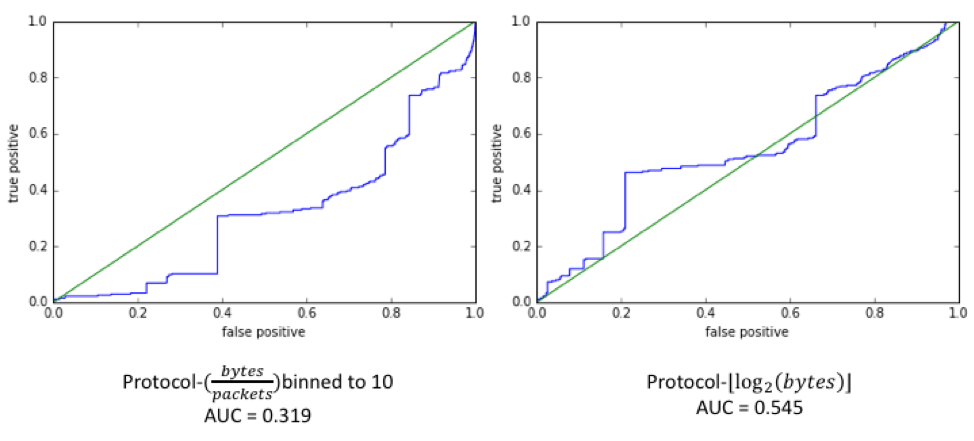}
\caption{The effect of tokenization on the performance of the PST analytic}
\label{fig6}
\end{figure}

This section presents experimental results of the PST approach to cyber anomaly
detection on the ISCX dataset. Figure \ref{fig6} shows the results for two
different types of tokens and their respective ROC curves. We experimented with two main
types of tokenization including proto-density binned in buckets of 10 (left) and proto-bytes binned
using log2 (right). For the ISCX dataset, using proto-bytes as a feature
significantly outperformed using proto-density.

\begin{figure}[h!]
\centering
\includegraphics[width=.95\columnwidth]{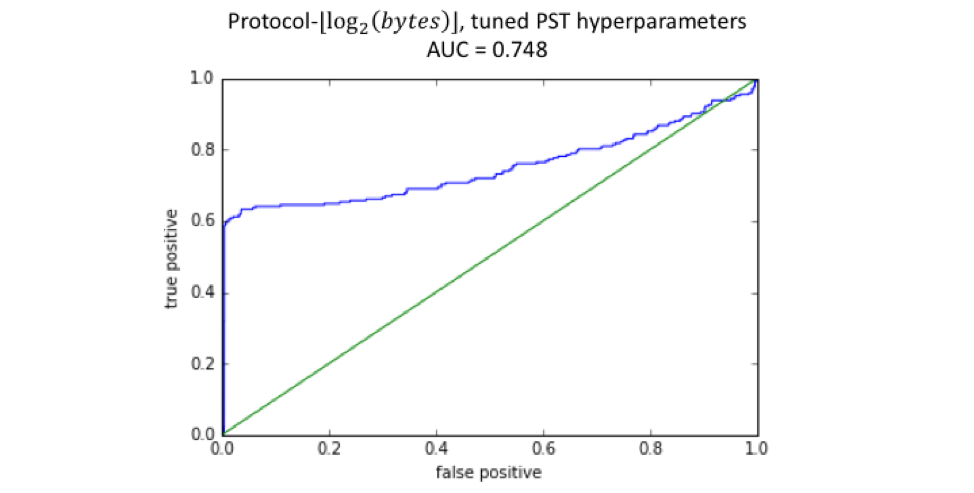}
\caption{The effect of tuning PST hyperparameters to increase the AUC}
\label{fig7}
\end{figure}

Another factor in PST performance is the tuning of the analytic hyperparameters.
As implemented, the PST has several hyperparameters. The tree depth specifies
the maximum depth of the model generated while the probability threshold value
is used to determine if a sequence is significant and is a candidate to add to
the PST. Raising the probability threshold makes the PST more restrictive. The
other parameters (tau, epsilon, and probability minimum) are essentially used
together to remove useless nodes from the PST model and as a smoothing factor.
Figure \ref{fig7} illustrates the effect of tuning these PST hyperparameters to
a specific dataset.

By tuning the PST hyperparameters, we are able to increase the AUC from 0.545
(shown on the right side of Figure \ref{fig6}) to 0.748. It should also be
noted that the shape of the ROC curve is of interest from the view of a cyber analyst. By
noting the sharp rise in the ROC curve at the beginning, we observe that the
results presented to a cyber analyst (assuming this same ordering) are less
likely to be false positives (i.e., less likely to degrade trust in the system).
In an operational environment, we would typically not have ground truth labels
for our data. It is important to build trust in the system by presenting minimal false positives to the cyber security analyst.

\section{Conclusions and Future Work}

This work demonstrates that there is a method to view and interpret cyber communications as a
language and that applying language-based analytic techniques to this new
synthetic language has potential. We have shown that there is value in viewing
network traffic as a language, and that PSTs can be used to characterize that language. By selectively engineering the
input features and tuning the PST hyperparameters, we can substantially increase
the AUC while maintaining a favorable ROC curve shape.

Future work in this area includes how to best retain the labels when aggregating
flows into sessions. While we have characterized on attack vs. non-attack
(normal), there are various types of attacks that exist and generating more nuanced labels than these binary indicators would be useful. Additional research
and experimentation should be devoted to the correct evaluation criteria for
results. Do cyber analysts care about the overall AUC or, perhaps more
importantly, the true positive rate for the first $n$ predicted anomalies/attacks?
In addition, experiments that evaluate how generalizable the results from
applying this methodology on the ISCX data are necessary. One method to do this
is to use additional labeled datasets from ISCX and then generate model-fit
comparisons on labeled and unlabeled data to show correlation.

\bibliographystyle{unsrt}
\bibliography{BDRRef_no_order}

\begin{thebibliography}{10}

\bibitem{gu2015novel}
Xiaodan Gu, Ming Yang, Jiaxuan Fei, Zhen Ling, and Junzhou Luo.
\newblock A novel behavior-based tracking attack for user identification.
\newblock In {\em Advanced Cloud and Big Data, 2015 Third International
  Conference on}, pages 227--233. IEEE, 2015.

\bibitem{sharafuddin2010know}
Esam Sharafuddin, Nan Jiang, Yu~Jin, and Zhi-Li Zhang.
\newblock Know your enemy, know yourself: Block-level network behavior
  profiling and tracking.
\newblock In {\em Global Telecommunications Conference (GLOBECOM 2010), 2010
  IEEE}, pages 1--6. IEEE, 2010.

\bibitem{verde2014no}
Nino~Vincenzo Verde, Giuseppe Ateniese, Emanuele Gabrielli, Luigi~Vincenzo
  Mancini, and Angelo Spognardi.
\newblock No nat'd user left behind: Fingerprinting users behind nat from
  netflow records alone.
\newblock In {\em Distributed Computing Systems (ICDCS), 2014 IEEE 34th
  International Conference on}, pages 218--227. IEEE, 2014.

\bibitem{abt2014small}
Sebastian Abt, Sebastian G{\"a}rtner, and Harald Baier.
\newblock A small data approach to identification of individuals on the
  transport layer using statistical behaviour templates.
\newblock In {\em Proceedings of the 7th International Conference on Security
  of Information and Networks}, page~25. ACM, 2014.

\bibitem{leung2005unsupervised}
Kingsly Leung and Christopher Leckie.
\newblock Unsupervised anomaly detection in network intrusion detection using
  clusters.
\newblock In {\em Proceedings of the Twenty-eighth Australasian conference on
  Computer Science-Volume 38}, pages 333--342. Australian Computer Society,
  Inc., 2005.

\bibitem{portnoy2001intrusion}
Leonid Portnoy, Eleazar Eskin, and Sal Stolfo.
\newblock Intrusion detection with unlabeled data using clustering.
\newblock In {\em In Proceedings of ACM CSS Workshop on Data Mining Applied to
  Security (DMSA-2001}. Citeseer, 2001.

\bibitem{munz2007traffic}
Gerhard M{\"u}nz, Sa~Li, and Georg Carle.
\newblock Traffic anomaly detection using k-means clustering.
\newblock In {\em GI/ITG Workshop MMBnet}, 2007.

\bibitem{bejerano2001variations}
Gill Bejerano and Golan Yona.
\newblock Variations on probabilistic suffix trees: statistical modeling and
  prediction of protein families.
\newblock {\em Bioinformatics}, 17(1):23--43, 2001.

\bibitem{liu2013incorporating}
Xumin Liu, Hua Liu, and Chen Ding.
\newblock Incorporating user behavior patterns to discover workflow models from
  event logs.
\newblock In {\em Web Services (ICWS), 2013 IEEE 20th International Conference
  on}, pages 171--178. IEEE, 2013.

\bibitem{ids2012}
Ali Shiravi, Hadi Shiravi, Mahbod Tavallaee, and Ali~A. Ghorbani.
\newblock Toward developing a systematic approach to generate benchmark
  datasets for intrusion detection.
\newblock {\em Computers and Security}, 31(3):357 -- 374, 2012.

\end{thebibliography}
\end{document}